\documentclass[aps,prb,twocolumn]{revtex4}

\usepackage{graphicx}
\graphicspath{ {FigsForPaper/} }

\usepackage{bm}
\usepackage{amsmath}
\usepackage{amssymb}

\newcommand{\ocite}{\onlinecite}

\newcommand{\iy}{\infty}

\newcommand{\pd}{\partial}

\newcommand{\dg}{\dagger}

\newcommand{\lt}{\left}
\newcommand{\rt}{\right}

\newcommand{\f}{\frac}
\newcommand{\tf}{\tfrac}

\newcommand{\sq}{\sqrt}
\newcommand{\lbl}{\label}

\newcommand{\eq}[1]{Eq.~(\ref{eq:#1})}
\newcommand{\eqs}[2]{Eqs.~(\ref{eq:#1}) and (\ref{eq:#2})}
\newcommand{\eqss}[3]{Eqs.~(\ref{eq:#1}), (\ref{eq:#2}), and (\ref{eq:#3})}

\newcommand{\eqn}[1]{(\ref{eq:#1})}
\newcommand{\eqsn}[2]{(\ref{eq:#1}) and (\ref{eq:#2})}

\newcommand{\figr}[1]{Fig.~\ref{fig:#1}}

\newcommand{\spc}{\mbox{ }}

\newcommand{\beq}{\begin{equation}}
\newcommand{\eeq}{\end{equation}}
\newcommand{\beqar}{\begin{eqnarray}}
\newcommand{\eeqar}{\end{eqnarray}}
\newcommand{\beqarn}{\begin{eqnarray*}}
\newcommand{\eeqarn}{\end{eqnarray*}}
\newcommand{\ba}{\begin{array}}
\newcommand{\ea}{\end{array}}
\newcommand{\bwt}{\begin{widetext}}
\newcommand{\ewt}{\end{widetext}}

\newcommand{\rarr}{\rightarrow}

\newcommand{\Z}{\mathbb{Z}}

\newcommand{\ix}{{\text i}}

\newcommand{\Hh}{\hat{H}}

\newcommand{\Sh}{\hat{S}}

\newcommand{\psih}{\hat{\psi}}

\newcommand{\Ec}{\mathcal{E}}

\newcommand{\Hc}{\mathcal{H}}

\newcommand{\Hch}{\hat{\Hc}}

\newcommand{\jb}{{\bf j}}

\newcommand{\pb}{{\bf p}}

\newcommand{\jbh}{\hat{\jb}}

\newcommand{\be}{\beta}

\newcommand{\De}{\Delta}

\newcommand{\tht}{\theta}
\newcommand{\Tht}{\Theta}
\newcommand{\ka}{\varkappa}

\newcommand{\e}{\epsilon}

\begin{document}
\title{
Paradoxical extension of the edge states across the topological phase transition
due to emergent approximate chiral symmetry
in a quantum anomalous Hall system
}

\author{Denis R. Candido$^{1,*}$, Maxim Kharitonov$^{2,*,\dg}$, J. Carlos Egues$^1$, Ewelina M. Hankiewicz$^2$}
\address{
{\normalfont $^*$These authors contributed equally\\ $^\dg$Corresponding author: maxim.kharitonov@physik.uni-wuerzburg.de}\\
\mbox{$^1$Instituto de F\'{\i}sica de S\~ao Carlos, Universidade de S\~ao Paulo, 13560-970, S\~ao Carlos, S\~ao Paulo, Brazil}\\
\mbox{$^2$Institute for Theoretical Physics and Astrophysics, University of W\"urzburg, 97074 W\"urzburg, Germany}
}

\begin{abstract}

We present a paradoxical finding that, in the vicinity of a topological phase transition in a quantum anomalous Hall system (Chern insulator),
topology nearly always (except when the system obeys charge-conjugation symmetry)
results in a significant extension of the edge-state structure beyond the minimal one required to satisfy the Chern numbers.
The effect arises from the universal gapless linear-in-momentum Hamiltonian of the nodal semimetal
describing the system right at the phase transition,
whose form is enforced by the change of the Chern number.
Its emergent approximate chiral symmetry results in an edge-state band
in the vicinity of the node, in the region of momenta where such form is dominant.
Upon opening the gap, this edge-state band is modified
in the gap region, becoming
``protected'' (connected to the valence bulk band with one end and conduction band with the other)
in the topologically nontrivial phase and
``nonprotected'' (connected to either the valence or conduction band with both ends)
in the trivial phase.
The edge-state band persists in the latter as long as the gap is small enough.

\end{abstract}

\maketitle

{\em Introduction and main result.}
In quantum anomalous Hall (QAH) systems (also known as Chern insulators)~\cite{Haldane1988,Liu2008,Chang2013,Check2014,Wang2014,Chang2015,Ryu2010,Hasan2010,Chiu2016},
the topological Chern number $C$ of an insulating phase
defines, via bulk-boundary correspondence~\cite{Ryu2010,Chiu2016},
the number of the edge-state bands that connect the valence and conduction bulk bands.
This is the only characteristic of the edge states required by quantum Hall (QH) topology.
Such states are {\em topologically protected} in the sense that
they cannot disappear under continuous deformations of the Hamiltonian without closing the gap in the bulk spectrum.

One could define {\em minimal edge-state structures} that are {\em sufficient} to satisfy a given Chern number.
In particular, in the topologically nontrivial (TnT) phase with $C=1$,
one topologically protected edge-state band,
having minimal extent in momentum space
just enough to connect the valence and conduction bands
(for more common band structures, this is typically the region of momenta dominated by the bulk gap), is sufficient.
In the topologically trivial (TT) phase with $C=0$, no edge states at all are required.
No edge states are also required
at the topological phase transition (TPT) between the TT and TnT phases, when the gap closes and the system is a semimetal,
since $C$ is not even well-defined there.

In principle, {\em topologically nonprotected} edge-state bands
that are connected to either the valence or conduction band with both ends could additionally exist.
Also, topologically protected edge-state bands could extend beyond the gap region.
Such additional edge-state structures
are not required by QH topology, but neither are they prohibited.

\begin{figure}[h!]
\includegraphics[width=.48\textwidth]{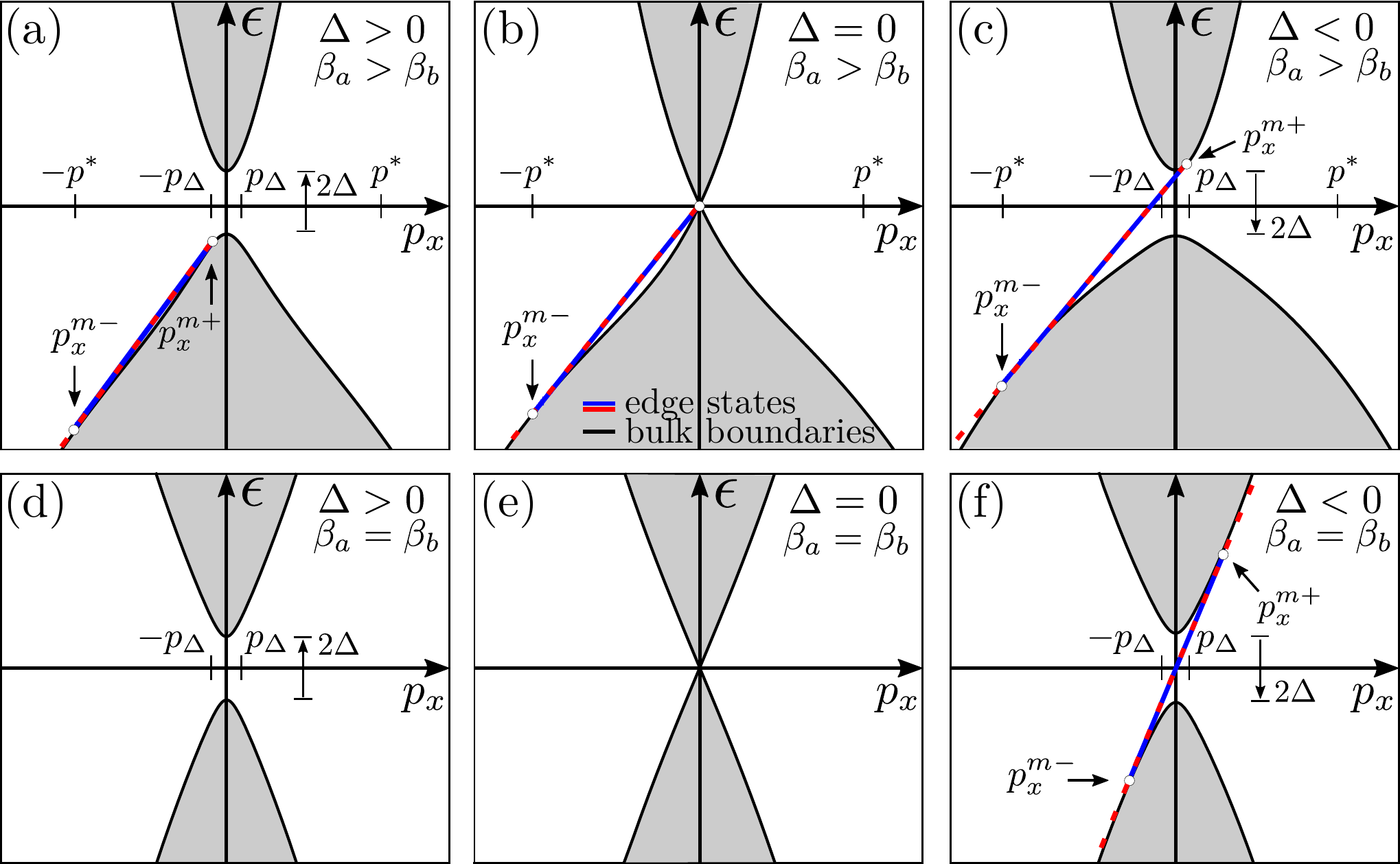}
\caption{
Edge states (blue) of a quadratic model \eqsn{H2}{bc2} of a QAH system
in the vicinity ($p_\De\ll p^*\sim\ka=v/\sq{\be_a\be_b}$) of the TPT.
At smaller momenta, the edge-state structure agrees with that (dashed red)
of the low-energy linear model [\eqss{H}{bc}{thtbe}, \figr{lin}].
Gray-shaded regions denote the continua of bulk states.
(a),(b),(c)
Due to the approximate chiral symmetry of the linear model, in the general case $\be_a\neq\be_b$
of unequal curvatures $\be_{a,b}$ of the conductance and valence bands,
the edge-state structure is extended
beyond the minimal one required to satisfy the Chern numbers.
(d),(e),(f) Only in the exceptional case $\be_a=\be_b$, when the system obeys charge-conjugation symmetry,
is the minimal edge-state structure realized.
Graphs (a),(b),(c) are plotted for $\be_a/\be_b=10$.
Graphs (a),(c),(d),(f) are plotted for $\ka/p_\De=3\sq{10}$.
}
\lbl{fig:quad}
\end{figure}

In this Rapid Communication, we present a paradoxical finding that, in the vicinity of a TPT in QAH system,
QH topology nearly always results in a significant extension of the edge-state structure beyond
the minimal one required to satisfy the Chern numbers, described above.
This generic behavior is illustrated in \figr{quad} with the quadratic model,
to be presented below, which describes one block of the Bernevig-Hughes-Zhang (BHZ)
model~\cite{BHZ} of a quantum spin Hall (QSH) system~\cite{KaneMele,BHZ,Koenig}.
This behavior has previously been noticed~\cite{JLi} in the BHZ model, but remained unexplained.

Applying the ideas recently formulated in Ref.~\ocite{KMH},
we show that this extension of the edge-state structure originates from the {\em emergent approximate chiral symmetry}
of the universal gapless linear-in-momentum low-energy Hamiltonian of the nodal semimetal describing the system right at the TPT,
which in its simplest form reads
\beq
    \Hh_0(\pb)=v(\tau_x p_x+\tau_y p_y),
\lbl{eq:H0}
\eeq
where $\tau_{x,y}$ are Pauli matrices in the space defined below
and $v$ is the velocity of its linear spectrum $\pm v|\pb|$, $\pb=(p_x,p_y)$.
This results in an edge-state band
that exists [\figr{quad}(b)] at the TPT on one side of the node (at least) in the region $|p_x|\lesssim p^*$
of momenta $p_x$ along the $y=0$ edge,
where such {\em asymptotic form} of the Hamiltonian is dominant;
$p^*$ is the scale, where higher-order-in-momentum terms in the asymptotic expansion
of the full Hamiltonian become comparable to the linear ones.
Upon opening a small gap $\De$, such that $p_\De=|\De|/v\ll p^*$,
the edge-state band is modified only in the smaller
region $|p_x|\lesssim p_\De$ dominated by the gap,
in accord with QH topology.
In the TT phase [$C=0$, \figr{quad}(a)], the edge-state band
exists in the region $p_\De\lesssim|p_x|\lesssim p^*$
and remains topologically nonprotected,
connected to the valence (in the case of the quadratic model presented in \figr{quad}) band with both ends
without crossing the gap.
In the TnT phase [$C=1$, \figr{quad}(c)],
the edge-state band becomes topologically protected, connected to the valence band with one end and conduction band with the other,
crossing the gap in the region $|p_x|\lesssim p_\De$,
but also extending well beyond it to the region $p_\De\lesssim|p_x|\lesssim p^*$.

Even though this extension of the edge-state structure is not topologically protected,
it is {\em nearly always} present, i.e., for all forms of the low-energy model,
except when it obeys charge-conjugation symmetry [\figr{quad}(d),(e),(f)],
which is an exceptional case that can likely be achieved only by accident or fine-tuning.
Only in this case, the edge-state structure is minimal in the sense described above.

Our finding is quite paradoxical
because the form \eqn{H0} of the nodal semimetal Hamiltonian at the TPT
is ultimately {\em enforced by QH topology}, to ensure the change of the Chern number across the TPT.
This form, due to its approximate chiral symmetry,
generates the extension of the edge-state structure,
which is, however, not required to satisfy the Chern numbers in either the TT ($C=0$) or TnT ($C=1$) phase.

{\em Low-energy Hamiltonian in the vicinity of a topological phase transition.}
We consider a QAH system~\cite{Haldane1988}, i.e.,
a two-dimensional (2D) band insulator with broken time-reversal symmetry, but no orbital effect of the magnetic field.
Such system belongs to class A of the general classification scheme~\cite{Ryu2010,Chiu2016} of topological systems
and is characterized by an integer $\Z$ bulk topological invariant well known as the Chern number.

We consider the vicinity of a TPT in a QAH system.
At the phase transition, two electron states, to be denoted $a$ and $b$,
become degenerate at some point in the Brillouin zone,
making the system gapless.
The low-energy linear-in-momentum Hamiltonian
for the wave function $\psih=(\psi_a,\psi_b)^T$
in the subspace of these two states can be written in its simplest form as
\beq
    \Hh(\pb)=v(\tau_x p_x+\tau_y p_y)+\De\tau_z,
\lbl{eq:H}
\eeq
where $\tau_{x,y,z}$ are the Pauli matrices
and $\pb=(p_x,p_y)$ is the momentum {\em deviation} from the degeneracy point.
In general, several additional terms could be present.
In Supplemental Material (SM)~\cite{SM},
we demonstrate that our findings for the simplest Hamiltonian $\Hh(\pb)$ presented below
also hold for the most general form of the linear Hamiltonian.

\begin{figure}
\includegraphics[width=.47\textwidth]{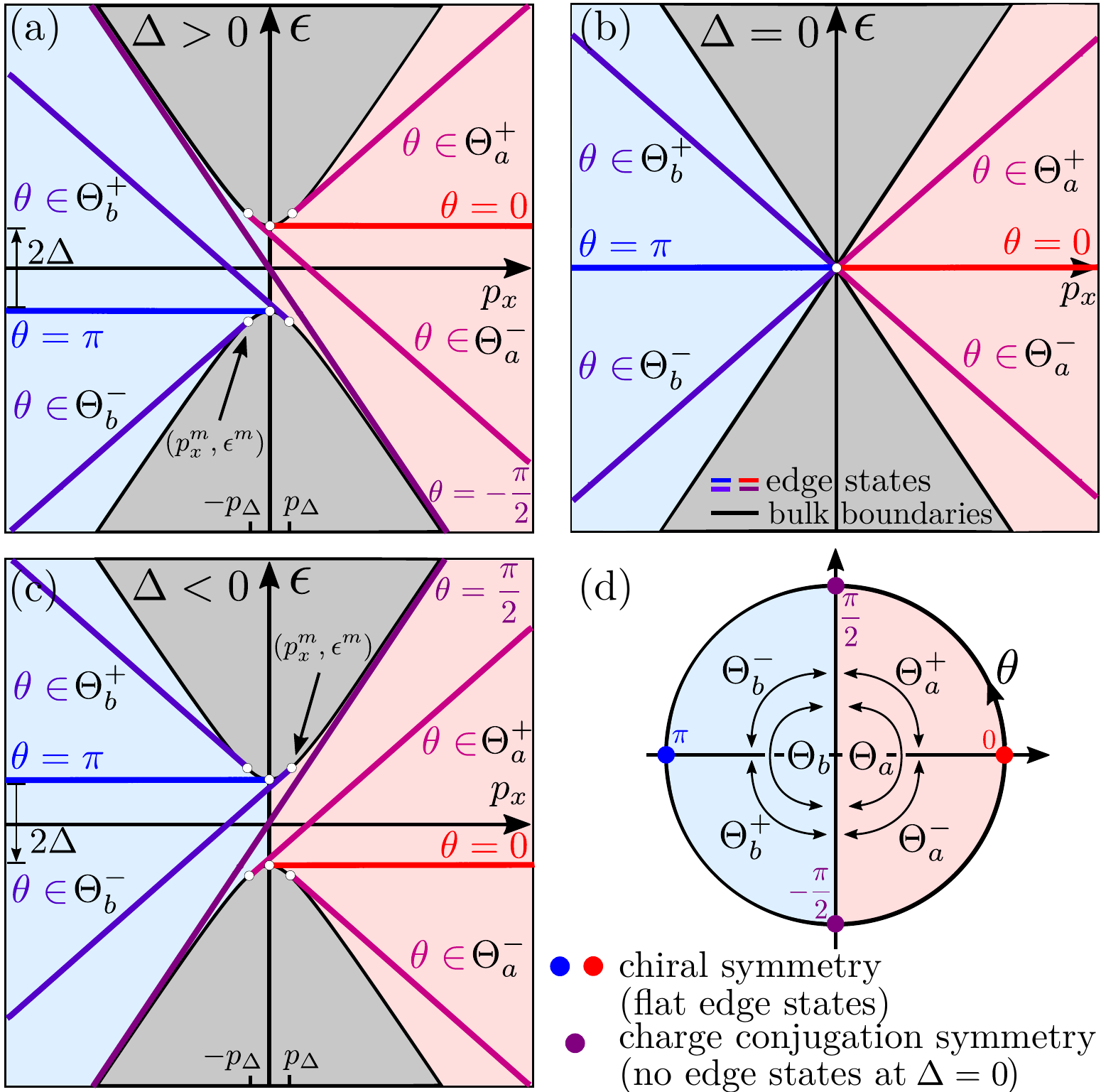}
\caption{
Generic edge-state structure [\eq{Ec}] in the vicinity of a TPT
in a QAH system, calculated for the simplest form \eqn{H} of the low-energy linear-in-momentum Hamiltonian
with the most general BC \eqn{bc}; see SM~\cite{SM} for the most general form of the Hamiltonian.
Due to the approximate chiral symmetry of the model,
for every value of the parameter $\tht$ of the BC \eqn{bc}
except $\pm\f\pi2$, when the system satisfies charge-conjugation symmetry,
the edge-state structure is extended
at the TPT [$\De=0$, (b)] and in both TT and TnT phases [$\De\gtrless0$, (a),(c)]
beyond the minimal one required to satisfy the Chern numbers.
In (a),(b),(c), gray-shaded regions denote the continua of bulk states.
(d) The circle of $\tht$. The stability regions $\Tht_a$ and $\Tht_b$ were subdivided
into the regions $\Tht_a^\pm$ and $\Tht_b^\pm$, respectively,
to indicate the location of the edge-state band in (a),(b),(c) for various
values of $\tht$.
}
\lbl{fig:lin}
\end{figure}

The two insulating QAH phases correspond to two signs of the gap $\De\gtrless 0$
and have a difference $C_{\De<0}-C_{\De>0}=1$ of the Chern numbers $C_{\De\gtrless0}$;
$\De=0$ is the TPT point, where the system is gapless nodal semimetal,
described by the Hamiltonian $\Hh_0(\pb)=\Hh(\pb)|_{\De=0}$ [\eq{H0}].
We assume that one of the phases is TT with a zero Chern number,
i.e., that possible full Chern numbers are $(C_{\De<0},C_{\De>0})=(1,0)$ or $(0,-1)$.

The Hamiltonian \eqn{H} is valid at momenta $p=|\pb|\lesssim p^*$
not exceeding the scale $p^*$ defined above.
Accordingly, the gap must also be small enough, so that $p_\De\lesssim p^*$.

{\em General boundary condition.}
Next, we supplement the low-energy Hamiltonian $\Hh(\pb)$ [\eq{H}]
with the boundary condition (BC) of the most general form,
describing termination of the sample, without any assumptions about the microscopic structure of the boundary.
Such form is constrained by the only fundamental requirement that the probability
current perpendicular to the boundary must vanish.
Such BC has recently been derived~\cite{Babak} in Ref.~\ocite{KMH}
for the gapless Hamiltonian $\Hh_0(\pb)$ [\eq{H0}].
Since the current operator $\jbh=\pd_\pb\Hh(\pb)=v(\tau_x,\tau_y)$ does not depend on $\De$ at all,
the BC remains exactly the same for finite $\De$.
Throughout this Rapid Communication, we consider the sample occupying the half-plane $y>0$.
The BC for the $y=0$ edge~\cite{axial} reads
\beq
    \psi_a(x,y=0)\sin\tf{\tht}2-\psi_b(x,y=0)\cos\tf{\tht}2=0.
\lbl{eq:bc}
\eeq
All unique forms of the BC are parameterized by the angle $\tht$ covering the {\em full circle},
shown in \figr{lin}(d).

{\em Generic edge-state structure.}
The edge states for the Hamiltonian \eqn{H} and BC \eqn{bc}
can be calculated analytically and are shown in \figr{lin}.
All possible forms are parameterized by the angle $\tht$
and represent the generic edge-state structure
in the vicinity of a TPT of QAH system, see also SM~\cite{SM}.
For every value of $\tht$ except $\pm\f{\pi}2$ (to be discussed separately below),
there exists {\em one} edge-state band
\beq
    \Ec(p_x)=v p_x\sin\tht+\De\cos\tht
\lbl{eq:Ec}
\eeq
for any value of the gap $\De$,
at $p_x>p_x^m(\tht)$  for $\tht\in\Tht_a=(-\f{\pi}2,\f{\pi}2)$
and at $p_x<p_x^m(\tht)$ for $\tht\in\Tht_b=(\f{\pi}2,\f{3\pi}2)$.
$(v p_x^m(\tht),\e^m(\tht))=\De(\tan\tht,1/\cos\tht)$
is the merging point of the edge-state dispersion relation \eqn{Ec}
with the boundaries $\pm\sq{(v p_x)^2+\De^2}$ of the continua of the bulk states.

Therefore, as the key property, the edge-state structure
in the vicinity of the TPT in a QAH system is nearly always (i.e., for all $\tht\neq\pm\f\pi2$)
extended beyond the minimal one required to satisfy the Chern numbers.
For the model \eqsn{H0}{bc} with $\De=0$ [\figr{lin}(b)],
these edge states have recently been found and studied in Ref.~\ocite{KMH}.
Applying the ideas expressed therein,
below we demonstrate that this extension of the edge states can be explained in terms of {\em chiral symmetry}.

{\em Edge states at the topological phase transition due to chiral symmetry.}
Chiral symmetry is one of the three main symmetries (along with time-reversal and charge-conjugation)
that give rise to topological behavior~\cite{Ryu2010,Chiu2016}.
A bulk Hamiltonian $\Hch(\pb)$ satisfies chiral symmetry,
if there exists a unitary operator $\Sh$ under which it changes its sign,
$\Sh\Hch(\pb)\Sh^\dg=-\Hch(\pb)$.

In 2D, chiral symmetry allows~\cite{Chiu2016}
for the existence of gapless topological semimetals
with flat edge-state bands at zero energy $\e=0$.
We notice that the gapless bulk Hamiltonian $\Hh_0(\pb)$ [\eq{H0}]
at the TPT obeys chiral symmetry with $\Sh=\tau_z$.
An important requirement for the topologically protected $\e=0$ edge states
to exist is that the system {\em with an edge} must respect chiral symmetry.
For a low-energy model, this means that not only the bulk Hamiltonian
but also the BC must respect chiral symmetry~\cite{KMH,SM}.
For the general BC \eqn{bc},
there are~\cite{KMH} only two discrete cases
\beq
    \psi_b(x,y=0)=0 \mbox{ or } \psi_a(x,y=0)=0
\lbl{eq:bcS}
\eeq
with $\tht=0$ or $\tht=\pi$, respectively, that obey chiral symmetry.

The model with the bulk Hamiltonian \eqn{H0} and one of the BCs \eqn{bcS}
represents a chiral-symmetric 2D topological semimetal.
For each of the chiral-symmetric BCs \eqn{bcS},
there is a flat edge-state band at $\e=0$
on one side of the node, for $p_x>0$ and $p_x<0$, respectively [\figr{lin}(b)],
ensured by the well-defined topological invariant of the node, the winding number 1.

The low-energy model of a QAH system in the vicinity of a TPT
with the Hamiltonian \eqn{H}
and BC \eqn{bc} does not generally
have exact chiral symmetry: in $\Hh(\pb)$,
the gap term $\De\tau_z$ breaks chiral symmetry $\Sh=\tau_z$
and the BC generally differs from one of the chiral-symmetric forms \eqn{bcS}.
However, as demonstrated in Ref.~\ocite{KMH},
exact chiral symmetry is not required for the edge states induced
by it to persist due to their {\em stability property}:
upon breaking chiral symmetry, the edge-state bands depart from $\e=0$, but continue to persist
as long as chiral-symmetric parts of the Hamiltonian and BCs are dominant.
Accordingly, the concept of a {\em stability region} was introduced therein,
as the region in the parameter space of chiral-asymmetric terms where the edge states
persist.

At the TPT $\De=0$ [\figr{lin}(b)],
the edge-state dispersion relation \eqn{Ec} reads~\cite{KMH} $\Ec(p_x)|_{\De=0}=v p_x \sin\tht$.
The circle of $\tht$ consists of two stability regions
$\Tht_a=(-\f\pi2,\f\pi2)$ and $\Tht_b=(\f\pi2,\f{3\pi}2)$,
for which the edge-state band is located at $p_x\gtrless0$, respectively.
The regions $\Tht_{a,b}$ contain the chiral-symmetric points $\tht=0$ and $\tht=\pi$, respectively,
corresponding to the BCs \eqn{bcS}~\cite{Thtab}.
As $\tht$ deviates from one of these points,
the edge-state band deviates from $\e=0$ acquiring a finite slope.
The edge-state band disappears by merging with the bulk bands only upon reaching $\tht=\pm\f\pi2$ points.
The stability regions $\Tht_{a,b}$ are thus separated only by two points $\tht=\pm\f\pi2$
and the edge states persist even for significant deviations from chiral symmetry.

To summarize this part, right at the TPT of a QAH system,
when the system is a nodal semimetal,
an edge-state band [\figr{lin}(b)]
nearly always [i.e., for all values of the parameter $\tht$ in the BC \eqn{bc}, except $\pm\f{\pi}2$]
exists (at least) in the region $|p_x|\lesssim p^*$,
where the linear terms of $\Hh_0(\pb)$ are dominant in the full Hamiltonian.
This band can be attributed to the approximate chiral symmetry of the model \eqsn{H0}{bc}.
This band is not required by QH topology
and is therefore an extension
of the minimal edge-state structure.

{\em Modification of the edge-state band by the gap.}
Introducing a small gap $\De$, such that $p_\De\ll p^*$,
modifies this edge-state band only in the smaller
region $|p_x|\lesssim p_\De$ dominated by the gap, in accord with QH topology.
Indeed, as Figs.~\ref{fig:lin}(a) and (c) show, in this region, the edge-state behavior is crucially sensitive to the sign of the gap,
i.e, to which QAH phase the system is in.
The numbers of the edge-state bands crossing the gap~\cite{mergecaveat}
with the sign of their velocities $\pd_{p_x}\Ec(p_x)=v\sin\tht$ are:
$0$ and $+1$ for $\De\gtrless 0$, respectively, if $\tht\in(0,\pi)$;
$-1$ and $0$ for $\De\gtrless 0$, respectively, if $\tht\in(\pi,2\pi)$.
The difference of these numbers for $\De\gtrless 0$ at each value of $\tht$
is in full accord with the difference $C_{\De<0}-C_{\De>0}=1$ of the Chern numbers,
manifesting the bulk-boundary correspondence of a QAH system.

However, also for a finite gap,
in the region $p_\De\lesssim |p_x|\lesssim p^*$,
the edge states are still determined by the linear chiral-symmetric part $\Hh_0(\pb)$ [\eq{H0}]
of the Hamiltonian $\Hh(\pb)$ [\eq{H}], which is dominant there.
As seen from Figs.~\ref{fig:lin}(a),(b),(c),
the edge-state structure in the region $p_\De\lesssim |p_x|\lesssim p^*$
is the same regardless of the presence of the gap and its sign
and represents the extension beyond the minimal structure required to satisfy the Chern numbers.

{\em The cases $\tht=\pm\f\pi2$ of charge-conjugation symmetry.}
For $\tht=\pm\f\pi2$, there is an edge-state band \eqn{Ec}
only in one of the insulating phases $\De\lessgtr0$, respectively,
while there are no edge states at the TPT $\De=0$ and in the other insulating phase $\De\gtrless0$.
Only in these cases of $\tht$ is the minimal edge-state structure realized.
In SM~\cite{SM}, we demonstrate
that the bulk Hamiltonian $\Hh(\pb)$ [\eq{H}] with the gap
and the BC \eqn{bc} at $\tht=\pm\f\pi2$
satisfy charge-conjugation symmetry.
Therefore, for $\tht=\pm\f\pi2$, the system with an edge obeys charge-conjugation symmetry.
Thus, interestingly, the only cases of the general BC \eqn{bc},
in which the edge states at the TPT $\De=0$ [\figr{lin}(b)]
induced by chiral symmetry
are absent, are the ones that have charge-conjugation symmetry.
This can be understood as follows. For charge-conjugation symmetry,
the edge-state spectrum has to satisfy $\Ec(p_x)=-\Ec(-p_x)$,
which is incompatible with the property that
chiral symmetry induces only one edge-state band on either side ($p_x>0$ or $p_x<0$) of a {\em linear} node.

{\em Edge states of the quadratic model.}
Above, we have established the generic behavior of the edge states (\figr{lin}) of a QAH system
in the vicinity of a TPT using the universal
low-energy model with the linear-in-momentum Hamiltonian,
given by \eqs{H}{bc} in its simplest form
and in SM~\cite{SM} in its most general form.
Any ``full'' model of a QAH system with an explicitly defined
behavior at all momenta and properties of the boundary
will asymptotically be described by this low-energy model in the vicinity of the TPT.
The information about both the band structure of the bulk Hamiltonian away from the node
and properties of the boundary will be fully contained in the BC of the low-energy model.

We demonstrate this point explicitly by considering the following QAH Hamiltonian
\beq
    \Hh_2(\pb)=\lt(\ba{cc} \De+\be_a p^2 & v p_-\\ v p_+&-\De-\be_b p^2\ea\rt),
    \spc p_\pm=p_x\pm\ix p_y,
\lbl{eq:H2}
\eeq
with momenta taking all values up to infinity.
One can recognize this model as one block of the BHZ model~\cite{BHZ}.
In addition to $\Hh(\pb)$ [\eq{H}],
this Hamiltonian contains diagonal quadratic terms $\be_{a,b} p^2$ with the curvatures $\be_{a,b}>0$.
These terms fully specify the topology, making the Chern number $C$ well defined.
The phases $\De\gtrless 0$ are TT and TnT with $C=0,1$, respectively.
We consider ``hard-wall'' BCs~\cite{hardwallbcs}
\beq
    \psi_a(x,y=0)=0,\spc \psi_b(x,y=0)=0
\lbl{eq:bc2}
\eeq
for the Hamiltonian \eqn{H2}.

The linearized Hamiltonian $\Hh(\pb)$ is obtained from $\Hh_2(\pb)$
just by neglecting the quadratic terms.
The BC of the form \eqn{bc} for the linearized Hamiltonian
is derived in SM~\cite{SM}.
As a result, the quadratic model of \eqs{H2}{bc2}
is described asymptotically in the vicinity of the TPT
by the linear model of \eqs{H}{bc},
with the angle $\tht$ of the BC determined by the ratio of the curvatures as
\beq
    \tan\tf{\tht}2=\sq{\be_a/\be_b}.
\lbl{eq:thtbe}
\eeq
As $\be_a/\be_b$ spans $(0,+\iy)$,
the angle $\tht$ spans the half-circle $(0,\pi)$,
i.e., in this particular model, only half of the possible forms of the general BC \eqn{bc} can be realized.
The BC \eqn{bc} becomes chiral-symmetric [\eq{bcS}]
only asymptotically in the limits $\be_a/\be_b\rarr 0,+\iy$,
i.e., for strong particle-hole asymmetry.

The edge states for the quadratic model of \eqs{H2}{bc2}
can actually be found analytically.
It turns out that its {\em exact}
edge-state spectrum is given~\cite{exactcaveat} by that \eqn{Ec} of the linear model with $\tht$ given by \eq{thtbe}.
For $\be_a>\be_b$, the exact edge-state band, shown in \figr{quad},
exists in the interval $p_x^{m-}<p_x<p_x^{m+}$, where
\beq
    p_x^{m\pm}=\f{\ka}2\f{\be_a-\be_b}{\be_a+\be_b}\lt(-1\pm \sq{1-8\f\De{v^2}\f{\be_a\be_b(\be_a+\be_b)}{(\be_a-\be_b)^2}}\rt)
\lbl{eq:pm}
\eeq
with $\ka=v/\sq{\be_a\be_b}$
are the merging points with the bulk spectrum of \eq{H2}.
The merging points $p_x^{m-}$ at larger momenta
can be used to define the scale $p^*\equiv |p_x^{m-}|$ of validity of the linear model.
For $p_\De\ll\ka$ and not too small $\be_a-\be_b$,
$
    p_x^{m-}\approx-\ka\f{\be_a-\be_b}{\be_a+\be_b}\sim \ka.
$

We see that, indeed, for a small gap ($p_\De\ll p^*$),
the exact edge-state behavior of the quadratic model at smaller momenta $|p_x|\ll p^*$
is in full accord with that of the linear model.
For any value $\be_a/\be_b\neq 1$,
the edge-state structure of the quadratic model in the vicinity of the TPT is extended beyond the minimal one
[Figs.~\ref{fig:quad}(a),(b),(c)].
Additionally, the well-defined merging point $p_x^{m-}$ at larger momenta
explicitly shows that the edge-state behavior is in agreement with QH topology.
In the TnT [\figr{quad}(c)] and TT [\figr{quad}(a)] phase,
the edge-state band is topologically protected and nonprotected, respectively.
Upon increasing the gap in the TT phase,
the merging points $p_x^{m\pm}$ come closer to each other
and the edge-state band eventually disappears at $p_\De\sim\ka$.
For $\be_a=\be_b$ [Figs.~\ref{fig:quad}(d),(e),(f)],
the quadratic model obeys charge-conjugation symmetry~\cite{SM}
and the edge states exist only in the TnT phase in the gap region,
realizing the minimal edge-state structure.

{\em Relation to real systems.}
Having demonstrated in SM the extension effect of the edge-state structure
for the most general low-energy model,
we expect it to be widespread in real QAH systems~\cite{Liu2008,Chang2013,Check2014,Wang2014,Chang2015,Heremans},
and also in QSH systems, at least when coupling between the Kramers blocks is weak~\cite{QSHfootnote}.
In SM, we demonstrate that the effect is quite pronounced for the parameters
of the BHZ model describing HgTe quantum wells.

{\em Acknowledgements.}
M. K. and E. M. H. were supported by the German Science Foundation (DFG) via Grant No. SFB 1170 ``ToCoTronics''
and the ENB Graduate School on Topological Insulators.
D. R. C. and J. C. E. were supported by CNPq, Capes, FAPESP, PRPUSP/Q-NANO,
FAPESP-BAYLAT  2016/50080-9 and acknowledge useful discussions with P. H. Penteado.


\begin{thebibliography}{99}
\bibitem{axial}
Due to axial rotation symmetry of the Hamiltonian $\Hh(\pb)$ [\eq{H}],
the edge-state spectrum will be the same for any orientation of the edge.
\bibitem{Thtab}
The regions $\Tht_{a,b}$ are labelled according to the dominant character of the wave function,
$|\psi_a|\gtrless|\psi_b|$, respectively.
\bibitem{mergecaveat}
We assume that the edge states eventually merge outside of the region $|p_x|\lesssim p^*$
of validity of the linear model with those bulk bands to which they are closest,
as is the case, e.g., for the quadratic model \eqn{H2} in \figr{quad}.
\bibitem{QSHfootnote}
If a QSH system possesses additional reflection symmetry along the direction perpendicular to the quasi-2D sample,
such as the original BHZ model~\cite{BHZ},
then it decouples into two QAH systems and our results apply directly.
If there is no exact reflection symmetry,
our results should still apply
at least when the effect of reflection-symmetry breaking is weak.
Generalization of our approach to stronger coupling should be possible.
\end{thebibliography}
\end{document}


\title{Supplemental Material:
Paradoxical extension of the edge states across the topological phase transition
due to emergent approximate chiral symmetry
in a quantum anomalous Hall system
}

\author{Maxim Kharitonov$^1$, Denis R. Candido$^2$, J. Carlos Egues$^2$, Ewelina M. Hankiewicz$^1$}
\address{
$^1$Institute for Theoretical Physics and Astrophysics, University of W\"urzburg, 97074 W\"urzburg, Germany\\
$^2$Instituto de F\'{\i}sica de S\~ao Carlos, Universidade de S\~ao Paulo, 13560-970, S\~ao Carlos, S\~ao Paulo, Brazil
}

\maketitle

\setcounter{equation}{0}
\renewcommand\theequation{S\arabic{equation}}

\setcounter{figure}{0}
\renewcommand\thefigure{S\arabic{figure}}

\setcounter{table}{0}
\renewcommand\thetable{S\arabic{table}}

\section{Most general linear model}

Here, we demonstrate that the main claim of the paper,
that the edge-state structure in the vicinity of the TPT in a QAH system
is nearly always significantly extended beyond the minimal one required to satisfy the Chern numbers,
is valid not only for the simplest form
(2)
of the linear Hamiltonian,
but also for its most general form.

\subsection{General bulk Hamiltonian, convenient basis}

The most general 2D Hamiltonian up to the linear order in momentum
for a two-component wave function, describing the vicinity of a TPT of a QAH system with no assumed symmetries,
has the form
\beq
    \Hh(\pb)=\sum_{\al=0,x,y,z} \tau_\al(\eps_\al+v_{\al x} p_x+v_{\al y} p_y),
\lbl{eq:Hgeninit}
\eeq
with 4 real energy parameters $\eps_\al$
and 8 real velocity parameters $v_{\al x}$ and $v_{\al y}$, i.e., with 12 real parameters,
out of which $\eps_0$ at $\tau_0$ is an inconsequential overall energy.
Here, $\tau_\al$ are the unity and Pauli matrices in the arbitrary initial basis of the two electron states
and $p_{x,y}$ are the cartesian momentum components in the arbitrary initial coordinate basis~\cite{xyzcaveat}.
Additionally, the orientation of the edge relative to the initial coordinate axes
is described by one angle.
And so, the edge-state problem is described by 12 real parameters:
11 relevant parameters of the bulk Hamiltonian and the angle specifying the orientation of the edge.

There is freedom in the independent choice of the wave-function $\psih=(\psi_a,\psi_b)^T$ and momentum $\pb=(p_x,p_y)$ bases.
The wave-function basis has 3 real parameters of the SU(2) transformation matrix $\Uh$,
\[
    \psih\rarr\Uh\psih.
\]
The momentum basis has one angle parameterizing the rotation matrix $\Rh$
and 2 parameters $\pb_0=(p_{x0},p_{y0})$ describing the choice of the origin,
\[
    \pb\rarr \Rh\pb+\pb_0.
\]
We exploit these 6 free real parameters to bring the most general Hamiltonian \eqn{Hgeninit}
to the most convenient form for the analysis of the BC and edge-state problem.

First, we use the angle of the rotation matrix $\Rh$ to orient the edge as $y=0$.
Next, we use 2 of the 3 parameters of the SU(2) matrix $\Uh$
to bring the matrix at $p_y$ momentum perpendicular to the edge $y=0$ to the diagonal form
(containing only $\tau_{0,z}$ and no $\tau_{x,y}$).
After this, the Hamiltonian in this new basis takes the form
\[
    \Hh(\pb)
    =\tau_0(v_{0x} p_x+v_{0y} p_y+\eps_0)
    +\tau_z(v_{zx} p_x+v_{zy} p_y+\eps_z)
    +\tau_x(v_{xx} p_x+\eps_x)
    +\tau_y(v_{yx} p_x+\eps_y).
\]
(In order not to overcomplicate the notation and since we are interested only in the final convenient form of the Hamiltonian,
we will preserve the same notation in all bases;
this should not lead to confusion.)
Next we use the remaining 3rd parameter of $\Uh$
to eliminate the velocity $v_{xx}$, which brings the Hamiltonian to the form
\[
    \Hh(\pb)
    =\tau_0(v_{0x} p_x+v_{0y} p_y+\eps_0)
    +\tau_y(v_{yx} p_x+ \eps_y)
    +\tau_z(v_{zx} p_x+v_{zy} p_y+\eps_z)
    +\tau_x \eps_x.
\]
Finally, by changing the origin $\pb_0$ of the momentum basis,
we eliminate two energies $\eps_{y,z}$.
Discarding also the inconsequential energy $\eps_0$ at $\tau_0$,
we arrive at the final desired form
\beq
    \Hh(\pb)
    =\tau_0(v_{0x} p_x+v_{0y} p_y)
    +\tau_y v_{yx} p_x
    +\tau_z (v_{zx} p_x+v_{zy} p_y)
    +\tau_x \De
\lbl{eq:Hgen}
\eeq
of the Hamiltonian, where we denoted $\eps_x\equiv\De$.
It is characterized by 6 real parameters: 5 velocities and 1 energy $\De$.
The bulk spectrum of \eq{Hgen} reads
\beq
    \eps_\pm(\pb)=v_{0y} p_y+v_{0x} p_x
    \pm\sq{(v_{zy} p_y+v_{zx} p_x)^2+(v_{yx} p_x)^2+\De^2}.
\lbl{eq:epsp}
\eeq

Counting the degrees of freedom,
we have used 6 available free parameters
to represent the edge-state problem with the most general Hamiltonian \eqn{Hgeninit} with 11 substantial parameters
and arbitrary edge orientation to that with the Hamiltonian \eqn{Hgen} with 6 parameters and $y=0$ edge.

The simplest Hamiltonian
\beq
    \Hh(\pb)
    \rarr v(\tau_y p_x+\tau_z p_y)
    +\tau_x \De,
\lbl{eq:Hmin}
\eeq
equivalent to Eq.~(2) via an additional change of basis,
is obtained from \eq{Hgen} when $v_{0y},v_{0x},v_{zx}=0$ and $v_{zy}=v_{xx}=v$.

\subsection{General boundary condition}

We now derive the most general form of the BC for the Hamiltonian \eqn{Hgen}.
Mathematically, such BC is a single linear homogeneous relation
between the two components $\psi_{a,b}(x,y=0)$ of the wave function at the edge $y=0$;
without loss of generality, it can be written as
\beq
    \ex^{\ix\f{\phi}2}\sin\tf\tht2\psi_a(x,y=0)-\ex^{-\ix\f{\phi}2}\cos\tf\tht2\psi_b(x,y=0)=0,
\lbl{eq:bcgeninit}
\eeq
parameterized by two real angles $\tht$ and $\phi$.

The BC must only satisfy the fundamental constraint that the current perpendicular to the edge vanishes
at the edge. For the Hamiltonian \eqn{Hgen}, the operator of the current perpendicular to the $y=0$ edge reads
\[
    \jh_y=\pd_{p_y}\Hh(p_x,p_y)=\tau_z v_{zy} +\tau_0 v_{0y}.
\]
It is given by the matrix at $p_y$ in \eq{Hgen},
which has conveniently been made diagonal.

Demanding that for any wave function satisfying \eq{bcgeninit} the current
$\psih^\dg(x,y=0)\jh_y\psih(x,y=0)=0$ vanish, we obtain
that the angle $\tht\equiv\tht_v\in(0;\pi)$ in \eq{bcgeninit}
is fixed by the velocity parameters of the bulk Hamiltonian as
\beq
    \cos\tht_v=-\f{v_{0y}}{v_{zy}},
\lbl{eq:thtv}
\eeq
while the angle $\phi$ may be arbitrary~\cite{singlenodecaveat}.

And so, the most general form of the BC for the $y=0$ edge
for the most general linear Hamiltonian represented in the form \eqn{Hgen}
reads
\beq
    \ex^{\ix\f{\phi}2}\sin\tf{\tht_v}2\psi_a(x,y=0)-\ex^{-\ix\f{\phi}2}\cos\tf{\tht_v}2\psi_b(x,y=0)=0.
\lbl{eq:bcgen}
\eeq
All unique forms of the BC are parameterized by the angle $\phi$ covering the {\em full circle},
similar to the angle $\tht$ in Eq.~(3).

\subsection{Edge states and chiral and charge-conjugation symmetries}

The edge states for the Hamiltonian \eqn{Hgen} and BC \eqn{bcgen} for the sample occupying the $y>0$ half plane
can be straightforwardly found.
For every value of $\phi$ except $\pm\f\pi2$, there exists one edge-state band
\beq
    \Ec(p_x)=\lt(v_{0x}+v_{zx}\cos\tht_v+v_{yx}\sin\tht_v\sin\phi\rt)p_x+\De\sin\tht_v \cos\phi
\lbl{eq:Ecgen}
\eeq
for any value of the gap $\De$, located at $p_x>p_x^m(\phi)$ for $\phi\in(-\f\pi2,\f\pi2)$
and at $p_x<p_x^m(\phi)$ and for $\phi\in(\f\pi2,\f{3\pi}2)$.
Here,
\[
    p_x^m(\phi)=\f{\De}{v_{yx}}\tan\phi, \mbox{ }
    \e^m(\phi)=\f\De{v_{yx}\cos\phi}[v_{yx}\sin\tht_v+(v_{0x}+v_{zx}\cos\tht_v)\sin\phi]
\]
are the momentum and energy, at which the edge-state spectrum $\Ec(p_x)$ merges
with the boundaries
\[
    \eps_{+x}(p_x)=\min_{p_y}\eps_+(p_x,p_y),\spc \eps_{-x}(p_x)=\max_{p_y}\eps_-(p_x,p_y)
\]
of the continua of the bulk spectrum [\eq{epsp}], given by
\beq
    \eps_{\pm x}(p_x)
    =(v_{0x}+v_{zx}\cos\tht_v)p_x\pm \sq{(v_{yx}p_x)^2+\De^2} \sin\tht_v.
\lbl{eq:eboundary}
\eeq

Therefore, for the most general form of the linear model [\eqs{Hgen}{bcgen}],
the edge-state structure in the vicinity of the TPT in a QAH system
is nearly always (i.e., for all $\phi\neq\pm\f{\pi}2$)
extended beyond the minimal one required to satisfy the Chern numbers.
As with the simplest Hamiltonian (2), below we analyze and explain
this behavior in terms of chiral and charge-conjugation symmetries, also providing some additional details.

A bulk Hamiltonian satisfies chiral symmetry~\cite{Ryu2010,Chiu2016}, if there exists a {\em unitary} operation
\beq
    \Sc[\psih](\rb)=\Sh\psih(\rb)
\lbl{eq:Spsi}
\eeq
on the wave function $\psih(\rb)=(\psi_a(\rb),\psi_b(\rb))^T$,
where $\Sh$ is a unitary matrix and $\rb=(x,y)$ is a radius vector,
under which the Hamiltonian changes its sign;
in momentum space, for $\psih(\pb)=\int\dx\rb\,\ex^{-\ix\pb\rb}\psih(\rb)$,
this reads
\beq
    \Sc[\Hc](\pb)=\Sh\Hch(\pb)\Sh^\dg=-\Hch(\pb).
\lbl{eq:SdefSM}
\eeq
We notice that the part
\beq
    \Hh_0(\pb)
    =\tau_y v_{yx} p_x
    +\tau_z (v_{zx} p_x+v_{zy} p_y)
\lbl{eq:H0gen}
\eeq
of the Hamiltonian $\Hh(\pb)$ [\eq{Hgen}] (with $v_{0x},v_{0y},\De=0$)
satisfies chiral symmetry with
\[
    \Sh=\tau_x,
\]
while the velocity $\tau_0(v_{0x} p_x+v_{0y} p_y)$ and gap $\tau_x\De$ terms
break such chiral symmetry.

Similarly, a bulk Hamiltonian satisfies charge-conjugation symmetry~\cite{Ryu2010,Chiu2016},
if there exists an {\em anti-unitary} operation
\[
    \Cc[\psih](\rb)=\Ch\psih^*(\rb)
\]
on the wave function, where $\Ch$ is a unitary matrix and $^*$ denotes complex conjugation,
under which the Hamiltonian changes its sign; in momentum space, this reads
\beq
    \Cc[\Hch](\pb)=\Ch\Hch^*(-\pb)\Ch^\dg=-\Hch(\pb).
\lbl{eq:Cdef}
\eeq
We notice that the {\em whole} Hamiltonian $\Hh(\pb)$ [\eq{Hgen}]
satisfies charge conjugation symmetry with
\[
    \Ch=\tau_z
\]
and $\Cc^2=+1$.

Next, we analyze the BC \eqn{bcgen} in terms of symmetries.
Chiral symmetry of the BC means that if a wave function $\psih(\rb)=(\psi_a(\rb),\psi_b(\rb))^T$ satisfies the BC,
then the wave function
$\Sc[\psih](\rb)=(\psi_b(\rb),\psi_a(\rb))^T$
transformed by the chiral symmetry operator \eqn{Spsi} also satisfies the same BC.
Substituting $\Sc[\psih](\rb)$ into the BC \eqn{bcgen}, we obtain
\[
    \ex^{+\ix\f{\phi}2}\sin\tf{\tht_v}2\psi_b(x,y=0)-\ex^{-\ix\f{\phi}2}\cos\tf{\tht_v}2\psi_a(x,y=0)=0.
\]
This condition is identical to \eq{bcgen}
only if $\tht_v=\f{\pi}2$, i.e., $v_{0y}=0$ [\eq{thtv}], and
\[
    \phi=0 \mbox{ or } \pi.
\]
Therefore,
\beq
    \psi_a(x,y=0)\mp\psi_b(x,y=0)=0,
\lbl{eq:bccs}
\eeq
respectively, are the only two chiral-symmetric forms of the BC \eqn{bcgen}, possible only when $v_{0y}=0$.

So, the chiral-symmetric Hamiltonian $\Hh_0(\pb)$ [\eq{H0gen}]
with one of the chiral-symmetric BCs \eqn{bccs} represents
a 2D chiral-symmetric semimetal with an edge.
Due to the well-defined winding number 1, such semimetal exhibits
flat edge-state bands that are topologically protected by chiral symmetry~\cite{Chiu2016}.
Indeed, it follows from \eq{Ecgen} that in this case the edge-state band $\Ec(p_x)\equiv0$
is flat and located at $p_x>0$ for $\phi=0$ and at $p_x<0$ for $\phi=\pi$.

The remaining gap $\tau_x\De$ and velocity $\tau_0(v_{0x}p_x+v_{0y}p_y)$ terms in the bulk Hamiltonian $\Hh(\pb)$ [\eq{Hgen}]
and the deviations of the angle $\phi$ of the BC \eqn{bcgen}
from $0$ or $\pi$ break chiral symmetry.
Nonetheless, the edge-state band \eqn{Ecgen} persist.
At the TPT $\De=0$, when the system is gapless,
the edge-state band exists for any values of $v_{0x}$ and $v_{0y}$
and any value of $\phi\neq\pm\f\pi2$: the velocity
\beq
    \Vc(\phi)=\pd_{p_x}\Ec(p_x)=v_{0x}+v_{xz}\cos\tht_v+v_{yx}\sin\tht_v\sin\phi
\lbl{eq:V}
\eeq
of the edge-state band \eqn{Ecgen} is always between the velocities
\beq
    v_{\pm x}=\pd_{p_x}\eps_{\pm x}(p_x)|_{\De=0}=v_{0x}+v_{xz}\cos\tht_v\pm|v_{yx}|\sin\tht_v
\lbl{eq:vr}
\eeq
of the dispersion relations \eqn{eboundary} determining the boundaries of the continua of the bulk spectrum,
$\Vc(\phi)\in[v_{-x},v_{+x}]$,
reaching them only at $\phi=\pm\f\pi2$, as shown in \figr{v}.

\begin{figure}
\includegraphics[width=.30\textwidth]{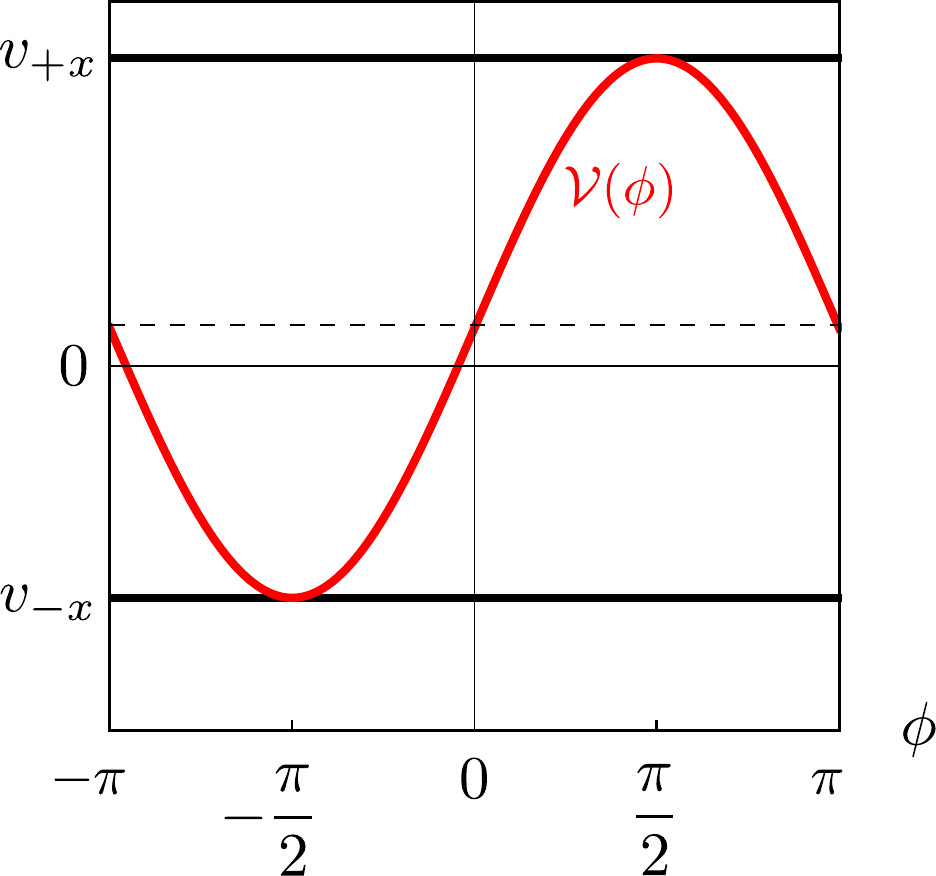}
\caption{
The velocity $\Vc(\phi)$ [\eq{V}]
of the edge-state band \eqn{Ecgen} at the TPT $\De=0$
of the most general linear Hamiltonian $\Hh(\pb)$ [\eq{Hgen}]
as a function of the angle parameter $\phi$ of the general BC \eqn{bcgen}.
The edge-state band exists
at $p_x>0$ for $\phi\in(-\f\pi2,\f\pi2)$
and at $p_x<0$ for $\phi\in(\f\pi2,\f{3\pi}2)$,
i.e., for every value of $\phi$ except $\pm\f\pi2$,
when the system obeys charge-conjugation symmetry.
At $\phi=\pm\f\pi2$,
the band merges with the boundaries of the continua of the bulk spectrum,
characterized by the velocities $v_{\pm x}$ [\eq{vr}].
This edge-state band at the TPT can be attributed to approximate chiral symmetry
of the linear model.
It is not required by QH topology
and thus represents an extension of the edge-state structure beyond the minimal one
required to satisfy the Chern numbers.
The graph is plotted for the velocity parameters
$(v_{0x},v_{0y},v_{zx},v_{zy})/v_{yx}
=(\f16,\f1{12},\f16,\f12)$.
}
\lbl{fig:v}
\end{figure}

The values $\phi=\pm\f{\pi}2$ are actually special symmetry-wise.
As with chiral symmetry, a BC is charge-conjugation-symmetric
if the transformed wave function $\Cc[\psih](\rb)=(\psi_a^*(\rb),-\psi_b^*(\rb))^T$
also satisfies the same BC. Substituting $\Cc[\psih](\rb)$ into the BC \eqn{bcgen},
we obtain
\[
        \ex^{-\ix\f{\phi}2}\sin\tf{\tht_v}2\psi_a(x,y=0)+\ex^{+\ix\f{\phi}2}\cos\tf{\tht_v}2\psi_b(x,y=0)=0,
\]
which is identical to \eq{bcgen} only if
\[
    \phi=\pm\f{\pi}2.
\]
Therefore,
\[
    \ex^{\pm\ix\f{\pi}4}\sin\tf{\tht_v}2\psi_a(x,y=0)-\ex^{\mp\ix\f{\pi}4}\cos\tf{\tht_v}2\psi_b(x,y=0)=0,
\]
respectively, are the only two charge-conjugation-symmetric forms of the BC \eqn{bcgen}.
Thus, as with the simplest Hamiltonian (2)~\cite{Cmin},
for the most general form \eqn{Hgen} of the linear Hamiltonian,
the only cases of the general BC \eqn{bcgen},
in which the edge-state band at the TPT $\De=0$ induced by chiral symmetry
is absent, are the ones that have charge-conjugation symmetry.
The intervals $\phi\in(-\f\pi2,\f\pi2)$ and $\phi\in(\f\pi2,\f{3\pi}2)$,
separated by the points $\phi=\pm\f{\pi}2$,
are the stability regions of the chiral-symmetric cases $\phi=0,\pi$.

The above-described edge-state band induced by chiral symmetry at the TPT
is not required by QH topology.
Therefore, it represents the extension of the edge-state structure
at TPT and also in gapped TnT and TT phases: a finite gap modifies this edge-state band
only in the momentum region dominated by the gap, as explained in the Main Text.

We have thus demonstrated with the most general form of the linear model [\eqs{Hgen}{bcgen}]
that the main claim of the paper
is completely general and not specific to the simplest form (1) of the bulk Hamiltonian considered in the Main Text:
the edge-state structure in the vicinity of the TPT in a QAH system
is nearly always significantly extended beyond the minimal one required to satisfy the Chern numbers.

\section{Quadratic model of a quantum anomalous Hall system}

Here, we present in more detail the properties
of the quadratic model described by Eqs.~(6) and (7).

\subsection{Boundary condition for the linear model \lbl{sec:quadbc}}

Here, we derive the BC of the form (3) for the linear model (2)
that asymptotically describes the quadratic model
with the Hamiltonian (6) and BCs (7) in the vicinity of its TPT $\De=0$.
The systematic procedure is as follows. Let us first find
the general solution to the Schr\"odinger equation
\[
    \Hh_2(p_x,\ph_y)\psih(y)=\e\psih(y)
\]
right at the TPT $\De=0$ for momentum $p_x=0$ and energy $\e=0$  exactly at the node
that contains no contributions growing into the bulk. We obtain
\[
    \psih(y)=\lt(\ba{c} \psi_a \\ \psi_b\ea\rt)+B\ex^{-\ka y}\lt(\ba{c} \sq{\be_b} \\ \sq{\be_a}\ea\rt),
\]
with the momentum scale
\beq
    \ka=\f{v}{\sq{\be_a\be_b}}
\lbl{eq:ka}
\eeq
and three independent {\em constant} coefficients $\psi_{a,b}$ and $B$.
Applying the BCs (7) to it and excluding $B$,
we arrive at the constraint
\beq
    \sq{\be_a}\psi_a-\sq{\be_b}\psi_b=0.
\lbl{eq:psibe}
\eeq

The quantity $\ka$ defines the validity scale of the low-energy model with the linear Hamiltonian (2):
the momentum scales of interest $p_x,\e/v,p_\De\ll \ka$ must be small.
Accordingly, the wave function $\psih(\rb)=(\psi_a(\rb),\psi_b(\rb))^T$ of the linear Hamiltonian (2)
must vary slowly at the spatial scale $1/\ka$.
To the leading order, the relation \eqn{psibe} will still apply
to the components $\psi_{a,b}\rarr\psi_{a,b}(x,y=0)$ of such wave function at the edge.
Thus, the relation \eqn{psibe} represent the BC of the form (3)
with the angle $\tht$ determined by the ratio of the curvatures as expressed in Eq.~(8).

\subsection{Charge-conjugation symmetry}

The quadratic Hamiltonian (6)
satisfies change-conjugation symmetry [\eq{Cdef}] with $\Ch=\tau_x$ only if $\be_a=\be_b$.
Its hard-wall BCs (7) also satisfy this symmetry.
Only in this case the minimal edge-state structure required to satisfy the Chern numbers is realized, shown in Fig.~1(d),(e),(f).

\subsection{Extension of the edge-state structure}

\begin{figure}
\includegraphics[width=.67\textwidth]{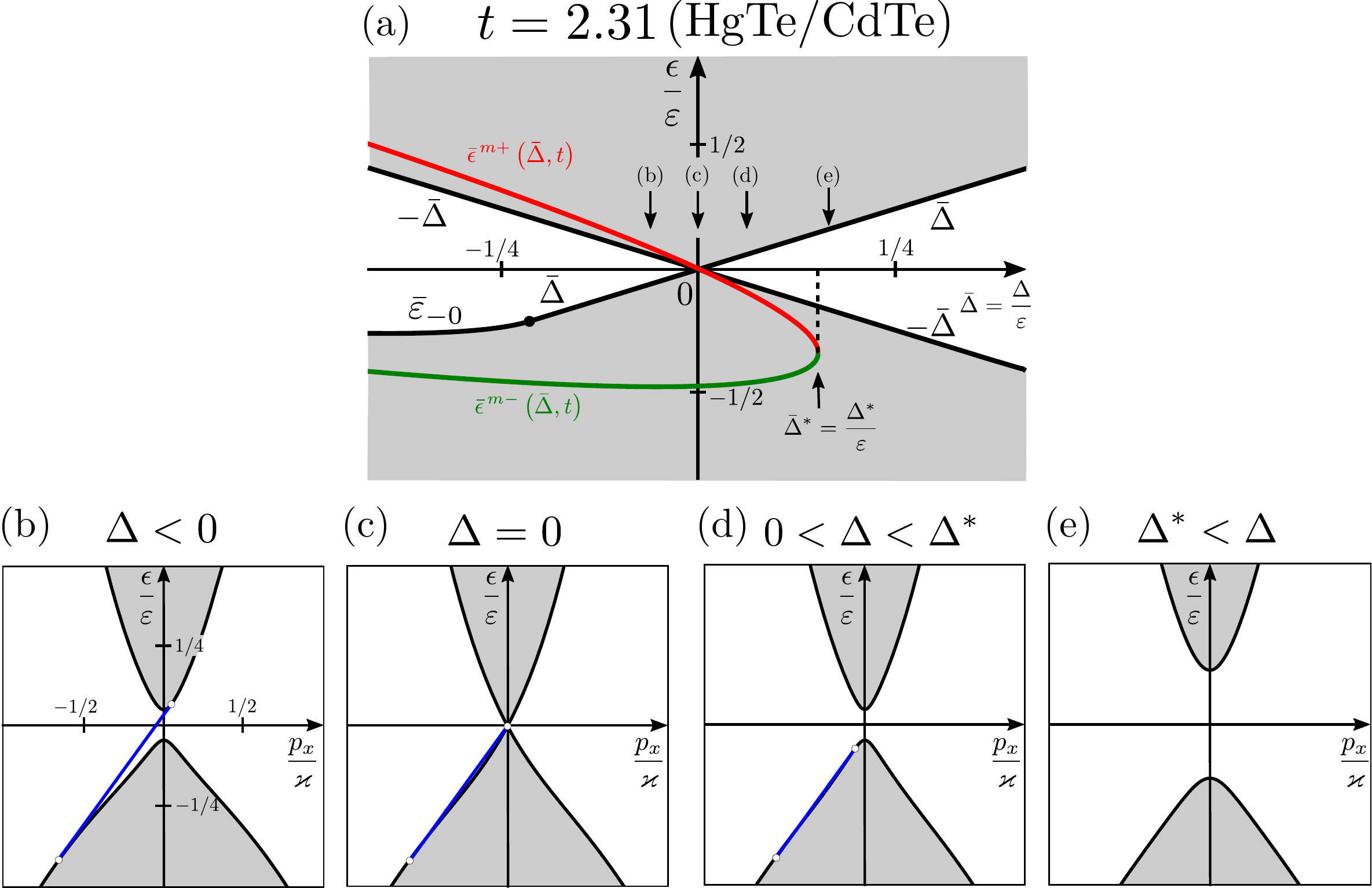}
\caption{
(a) The dependence on the dimensionless gap $\Der=\De/\eps$
of the dimensionless energies $\er^{m\pm}(\Der,t)=\e^{m\pm}/\eps$ [\eq{erm}, red and green]
of the merging points of the edge-state band in the quadratic model [Eqs.~(6) and (7)]
of a QAH system (also describing one block of the BHZ model of a QSH system).
The grey-shaded regions denote the continua of the bulk states of the conduction and valence bands.
(b)-(e) The samples of the edge-state spectra (blue)
at various values of $\Der$, indicated in (a), describing different regimes.
The graphs are plotted for the value $t=2.31$ [\eq{bet}]
that corresponds to the parameters (\tabr{quad})
estimated in Ref.~\ocite{book} for a HgTe quantum well at the topological phase transition.
For presentation purposes,
we used the plot aspect ratios different from 1, which can be deduced from the provided tick values.
}
\lbl{fig:quadHgTe}
\end{figure}

\begin{figure}
\includegraphics[width=.50\textwidth]{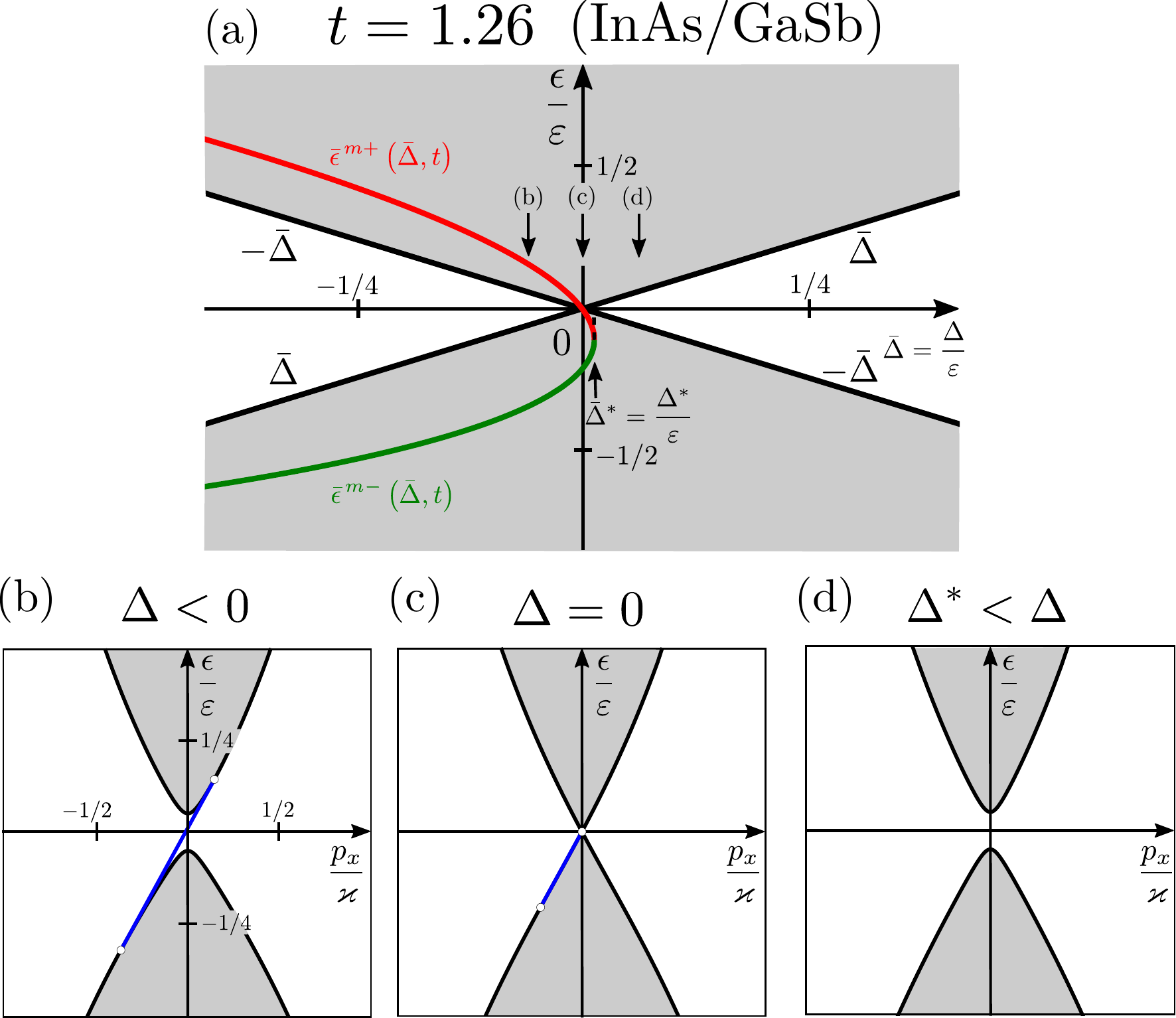}
\caption{
Same as in \figr{quadHgTe} for the value $t=1.26$ [\eq{bet}]
that corresponds to the parameters (\tabr{quad})
estimated in Ref.~\ocite{book} for an InAs quantum well at the topological phase transition.
}
\lbl{fig:quadInAs}
\end{figure}

Here, we discuss in more detail the effect of the extension of the edge-state structure
beyond the minimal one required to satisfy the Chern numbers
in the quadratic model of a QAH system described by Eqs.~(6) and (7).

The model is described by four parameters, the gap $\De$,
velocity $v$, and curvatures $\be_{a,b}$.
We observe that instead of $\be_{a,b}$
it is more physically insightful to consider the combinations
\beq
    \be=\sq{\be_a\be_b},
\spc
    t=\sq{\f{\be_a}{\be_b}}
\lbl{eq:bet}
\eeq
as independent parameters, in terms of which
\[
    \be_a=\be t,\spc \be_b=\be \f1t.
\]
This way, the geometric mean $\be$ sets
the overall momentum $\ka$ [\eq{ka}] and energy
\beq
    \eps=v\ka
    =\f{v^2}\be
\lbl{eq:eps}
\eeq
scales of the quadratic model, which, in particular, define the range of validity of the linear model, as explained in \secr{quadbc}.
The dimensionless parameter $t$, on the other hand, fully characterizes
the asymptotic edge-state structure [Eq.~(4)] in the vicinity of the TPT
through Eq.~(8) for the parameter of the BC (3) of the linear model.
This ``asymmetry'' parameter $t$ describes the deviation
from the charge-conjugation symmetry point $t=1$, at which the extension of the edge-state structure is absent.
The more $t$ deviates from 1, the closer one is to the chiral-symmetric case
and the flatter the edge-state band (4) is.
The quadratic model is thus fully characterized by two dimensionless parameters, $t$ and $\Der=\De/\eps$,
while the two dimensional parameters, say, $\ka$ and $\eps$,
set the natural units of measurement.

When measured in the units of $\ka$ and $\eps$,
the momenta $p^{m\pm}$ [Eq.~(9)] of the merging points of the edge-state band
and their energies
\beq
    \e^{m\pm}=\f1{(\be_a+\be_b)^2}
    \lt[-\De(\be_a^2-\be_b^2)+v^2(\be_a-\be_b)\lt(-1\pm\sq{1-\f{8\De\be_a\be_b(\be_a+\be_b)}{v^2(\be_a-\be_b)^2}}\rt)\rt]
\lbl{eq:em}
\eeq
are presentable as dimensionless functions
\[
    \pr_x^{m\pm}(\Der,t)=\f{p_x^{m\pm}}\ka
    =\f12\f{t-\f1t}{t+\f1t}\lt(-1\pm \sq{1-\f{\Der}{\Der^*}}\rt),
\]
\beq
    \er^{m\pm}(\Der,t)=\f{\e^{m\pm}}{\eps}
    =-\Der\f{t-\f1t}{t+\f1t}
    +\f{t-\f1t}{(t+\f1t)^2}\lt(-1\pm\sq{1-\f{\Der}{\Der^*}}\rt)
\lbl{eq:erm}
\eeq
of the dimensionless gap $\Der$ and parameter $t$.
Here,
\beq
    \De^*=\f{v^2(\be_a-\be_b)^2}{8\be_a\be_b(\be_a+\be_b)}
    =\eps \Der^*,\spc \Der^*=\f18\f{(t-\f1t)^2}{(t+\f1t)}
\lbl{eq:De*}
\eeq
is the critical value of the gap in the TT phase, at which the edge-state band disappears.
The edge-state band exists for all $\De<\De^*$, i.e., in the TnT phase for $\De<0$
(where, in the vicinity of the TPT, $|\De|\ll\eps$, it extends beyond the gap region),
at the TPT $\De=0$, and in the TT phase for $0<\De<\De^*$ up to this critical value.
As explained in the Main Text, this behavior constitutes
the extension of the edge-state structure beyond the minimal one required to satisfy the Chern numbers.
The critical gap $\De^*$ characterizes the magnitude of the extension effect and is
determined by both the deviation of $t$ from 1 (the case of charge-conjugation symmetry)
and the energy scale $\eps$ [\eq{eps}].

In \figsr{quadHgTe}{quadInAs}, we plot the dependence of the dimensionless energies $\er^{m\pm}(\Der,t)$ [\eq{erm}]
of the merging points on dimensionless gap $\Der$, for two values of $t$
that correspond to the sets of parameters, presented in \tabr{quad},
of the BHZ model that have been estimated in Ref.~\ocite{book}
for QSH systems realized in quantum wells.
We use the parameters for the cases that are very close to the TPT,
with thicknesses $d=6.1\x{nm}$ and $9.0\x{nm}$ for HgTe/CdTe and InAs/GaSb/AlSb quantum wells, respectively.
For HgTe, \figr{quadHgTe},
both the deviation of the asymmetry parameter $t=2.31$ from 1
and the energy scale $\eps=354\x{meV}$ are quite large,
giving an experimentally appreciable critical value $\De^*=57\x{meV}$
and implying quite a pronounced extension effect.
For InAs, on the other hand, \figr{quadInAs},
the asymmetry parameter $t=1.26$ is close to 1, i.e., the system is much closer to the case of charge-conjugation symmetry,
and the energy scale $\eps=5.04\x{meV}$ is small due to the much smaller velocity $v$ compared to the HgTe case;
as a result, the extension effect is weak, with an experimentally small critical value $\De^*=0.070\x{meV}$.

In \figsr{quadHgTe}{quadInAs},
the grey-shaded regions denote the continua of the conduction- and valence-band states
\[
    \eps_\pm(p)=\tf12(\be_a-\be_b)p^2\pm\sq{[\De+\tf12(\be_a+\be_b)p^2]^2+(vp)^2}, \spc p^2=p_x^2+p_y^2,
\]
of $\Hh_2(\pb)$ [Eq.~(6)], respectively, as all momenta $p$ are spanned.
These energy regions are bounded
by the minimum
\[
    \eps_{+0}=\min_{p}\eps_+(p)
\]
of the conduction band
and maximum
\[
    \eps_{-0}=\max_{p}\eps_-(p)
\]
of the valence band, given by
\[
    \eps_{\pm0}=\lt\{\ba{c}
        \pm|\De|,\spc \De_{\pm0}<\De,\\
        \f1{(\be_a+\be_b)^2}\lt[-\De(\be_a^2-\be_b^2)-v^2(\be_a-\be_b)\pm2v\sq{\be_a\be_b[-2\De(\be_a+\be_b)-v^2]}\rt], \spc \De<\De_{\pm0}, \ea\rt.
\]
\[
    \De_{+0}=-\f{v^2}{2\be_b},\spc
    \De_{-0}=-\f{v^2}{2\be_a}.
\]
In the dimensionless units,
\[
    \epsr_{\pm0}(\Der,t)=\f{\eps_{\pm0}}{\eps}
    =\lt\{\ba{c}
        \pm|\Der|,\spc \Der_{\pm0}<\Der,\\
        -\Der\f{t-\tf1t}{t+\f12}-\f{t-\tf1t}{(t+\tf1t)^2}\pm\f{2}{(t+\tf1t)^2}\sq{-2\Der(t+\tf1t)-1}, \spc \Der<\Der_{\pm0}, \ea\rt.
\]
\[
    \Der_{+0}=\f{\De_{+0}}\eps=-\f{t}{2},\spc
    \Der_{-0}=\f{\De_{-0}}\eps=-\f1{2t}.
\]
At $\De_{\pm0}<\De$, the dispersion relation $\eps_\pm(p)$ is monotonic in $p$ and its absolute minimum/maximum (for $\pm$, respectively)
$\eps_\pm(0)=\pm|\De|$ is reached at $p=0$.
However, at $\De<\De_{\pm0}$, $\eps_\pm(p)$
becomes nonmonotonic (``mexican-hat'' regime)
and its absolute minimum/maximum
is reached at a finite value of $p$.
We point out that the latter regime occurs
at large enough negative $\De<\De_{\pm0}$,
i.e., away from the TPT, where the extension effect we predict is not present anymore.
The formulas for the edge-state dispersion relation [Eqs.~(4) and (8)] and merging points [Eqs.~(9) and \eqn{em}]
are valid in all regimes of $\De$.
In \figr{quadHgTe} for HgTe, the transition point $\Der_{-0}$ is within the plot range,
while $\Der_{+0}$ is not; in \figr{quadInAs} for InAs, both $\Der_{\pm 0}$ are outside of the plot range.

For the BHZ model of a QSH system, the model (6) and (7) of the QAH system represents one Kramers block,
and the full spectrum is obtained by from Figs. 1 and S2
by adding the counterparts obtained by time reversal.

\begin{table}
\begin{tabular}{|c|c|c|c|c|c|c|c|}
\hline
& $v(\x{meV}\cd\x{nm})$ & $\be_a (\x{meV}\cd\x{nm}^2)$ & $\be_b (\x{meV}\cd\x{nm}^2)$ & $\be(\x{meV}\cd\x{nm}^2)$
& $t$ & $\eps(\x{meV})$ & $\De^*(\x{meV})$
\\\hline
HgTe, $d=6.1\x{nm}$ & 378& 931& 175& 404& 2.31 & 354 & 57\\
InAs, $d=9.0\,\x{nm}$ & 62& 963& 603& 762& 1.26 & 5.04 & 0.070
\\\hline
\end{tabular}
\caption{Parameters of the BHZ model estimated in Ref.~\ocite{book}
for HgTe and InAs quantum wells at the topological phase transition;
$d$ is the thickness of the active layer.
Similar values of parameters for a HgTe quantum well were estimated in Ref.~\ocite{Rothe}.
}
\lbl{tab:quad}
\end{table}